\documentclass[prd,reprint,nofootinbib,superscriptaddress,preprintnumbers,showpacs]{revtex4-1}
\usepackage{amsmath}	
\usepackage{amssymb}	
\usepackage{amstext}	
\usepackage{latexsym}	
\usepackage{bm}			
\usepackage{graphicx,color}	
\usepackage{slashed}	
\usepackage{makecell}	
\usepackage{multirow}	
\usepackage{caption}	
\usepackage{hyperref}	

\allowdisplaybreaks[4]
\pdfoutput=1

\begin{document}
\title{The Theoretical Study of $\pi^+p\to pa_0^+\phi$ Reaction}
\date{\today}
\author{Aojia Xu}
\affiliation{School of Physics, Dalian University of Technology, Dalian 116024, People's Republic of China}
\author{Ruitian Li}
\affiliation{School of Physics, Dalian University of Technology, Dalian 116024, People's Republic of China}
\author{Xuan Luo}
\affiliation{School of Physics and Optoelectronics Engineering, Anhui University, Hefei 230601, People's Republic of China}
\author{Hao Sun}
\email{haosun@dlut.edu.cn}
\affiliation{School of Physics, Dalian University of Technology, Dalian 116024, People's Republic of China}
\begin{abstract}
We study the production of hyperon resonances in the $\pi^+p\to pa_0^+\phi$ reaction within an effective Lagrangian approach. The model includes the production of $\Delta(1940)$ and $\Delta(1920)$ in the intermediate state excited by the $\rho$ and $p$ exchanges between the initial proton and $\pi$ meson. Due to the large contribution of $\rho$ exchange in $\Delta(1940)$ production, $\Delta(1940)$ is found a significant contribution near the threshold in this reaction. We provide total and differential cross section predictions for the reation and discuss the possible influence of model parameters, off-shell effects and branching ratios for $\Delta(1940)\to pa_0$ and $\Delta(1920)\to pa_0$ decays, which will be useful in future experimental studies. This reaction can provide a platform for studying the features of $\Delta(1940)$ resonance, especially the coupling to $pa_0$ channel.
\end{abstract}
\maketitle
\section{introduction}
The investigation of the meson-baryon interactions at low energies plays an important role in exploring the features of resonances, where the isospin 3/2 $\Delta^{++}(1232)$ excited states consisting of three identical quarks have been of great interest in hadron physics. However, experiments on $\Delta^*$ resonances are not as extensive as those on nucleon resonances. A pioneering measurement in the $\gamma p\to p\pi^0\eta$ reaction for the existence of a parity doublet of $\Delta^*$ resonances with total angular momentum $J=3/2$, $\Delta(1940)$ and $\Delta(1920)$, which can decay into a $pa_0(980)$ pair was first reported by CB-ELSA Collaboration~\cite{CB-ELSA:2007xbv}. Whereafter CB-ELSA Collaboration further measured the presence of six $\Delta^*$ resonances in $\gamma p\to p\pi^0\eta$ reaction~\cite{CB-ELSA:2008zkd}. After this, the researches on $\Delta^*$ resonances coupled with $pa_0(980)$ channel are rather few. On the one hand, on account of a small branching ratio for $\Delta^*\to pa_0$ decay, it is natural to think that $\Delta^*$ resonances has a small coupling with $pa_0(980)$ channel. But in fact, the $\Delta(1920)pa_0$ coupling constant is not small, even at a branching ratio of 1\%~\cite{Wang:2023lnb}. On the other hand, in recent years, most of the knowledge about $\Delta^*$ resonances came from photon induced reactions~\cite{CLAS:2023akb, CBELSATAPS:2021osa, CLAS:2018drk, CBELSATAPS:2015kka, CBELSATAPS:2014wvh, Kohri:2006yx}. By contrast, current $\pi N$ scattering data were obtained almost more than 30 years ago~\cite{Candlin:1982yv, Hart:1979jx, Winik:1977mm}, which typically have lower statistics and larger uncertainty. It is a fact that previous $\pi N$ scattering experiments concentrated on the single meson production processes, making research on multimeson production rather limited. However, multimeson production processes may also provide important information~\cite{Chen:2020szc, Fan:2019lwc, Xiao:2015zja, Lu:2014yba}. Beyond that, HADES collaboration have attempted to further generate excited $\Delta^{++}$ resonances from $NN$ collision~\cite{HADES:2011ryy}.

In the present work, we propose the $\pi^+p\to pa_0^+\phi$ reaction, which is a very ideal channel for investigating the $\Delta^*$ resonances with isospin 3/2 since the contributions of $N^*$ with isospin 1/2 are filtered out by charge conservation in this channel. In this reaction, we only need to take into account the coupling of the intermediate resonance with final $pa_0$ pair, without considering other channels, for which in the energy region under study no significant $\Delta^*$ resonance signals decay into a channel associated with $\phi$ meson are found~\cite{ParticleDataGroup:2022pth}. This reaction excludes other $\Delta^*$ resonances that are not coupled to $pa_0$ channel, making it a very clean process to study the coupling of $\Delta^*$ resonance with $pa_0$ pair. Moreover, the threshold energy of $pa_0$ channel is about 1.918~GeV, which is very close to the mass of $\Delta(1940)$ and $\Delta(1920)$ resonances. With these advantages, such a reaction is not only important for understanding the reaction mechanism itself, but also helpful for learning about the coupling of $\Delta^*$ resonances with the exchanged particles.

We investigate the reaction by using an effective Lagrangian approach, focusing on the production of $\Delta(1940)$ resonance. The approach of effective Lagrangian calculating the reaction cross section is widely used to investigate the process of particle collisions for exploring the reaction mechanism between initial and final particles~\cite{Xu:2024xso, Liu:2019ojr, Xie:2014kja, Wang:2014ofa, Oh:2007jd}. In our model, the $\Delta(1940)$ resonance is excited by $t-$channel $\rho$ exchange and $u-$channel $p$ exchange between the initial proton and $\pi$ meson, and $\Delta(1920)$ resonance will only be excited by $u-$channel $p$ exchange between the initial proton and $\pi$ meson, on acconut of its lacking of $\Delta(1920)N\rho$ coupling. Other meson or baryon exchanges are forbidden by the law of isospin conservation. Since the involved $\Delta(1940)$ resonance have relatively large decay branch ratios to $N\rho$ channel, it is natural to expect that the $t-$channel $\rho$ exchange play important roles for the excitation of $\Delta(1940)$ in this reaction. The predictions of the total cross section and angular distribution, as well as $pa_0$ invariant mass distribution are presented in our work, which will be helpful for future comparison with the experimental data. We also provide a discussion for the dependence of total and differential cross sections on model parameters and branching ratios for $\Delta(1940)\to pa_0$ and $\Delta(1920)\to pa_0$ decays.

Our work is organized as follows. In Sec.~II, we introduce the formalism and ingredients necessary of each amplitude in our model and obtain the concrete form of amplitudes. The numerical results of the total and differential cross sections for $\pi^+p\to pa_0^+\phi$ reaction are presented in Sec.~III. Finally, a short conclusion is made in Sec.~IV.

\section{Formalism}
\begin{figure}[htpb]
	\centering
	\includegraphics[width=0.4\textwidth]{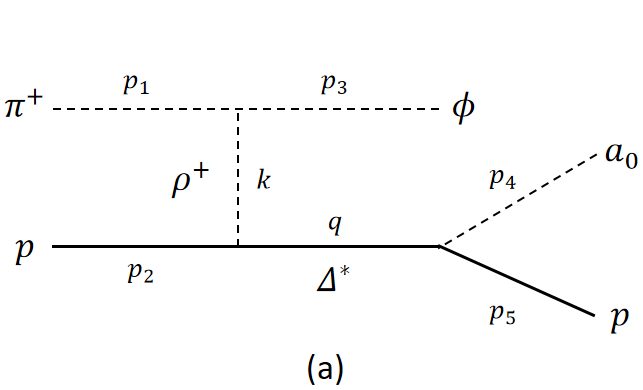}\hypertarget{1a}{}\\
	\includegraphics[width=0.4\textwidth]{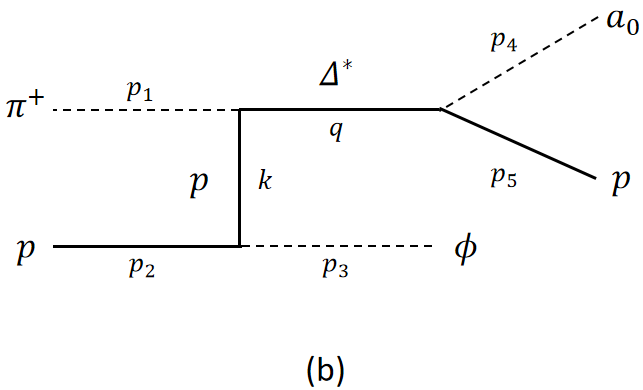}
	\captionsetup{justification=raggedright}
	\caption{(a) t- (b) u-channel exchanges Feynman diagrams for $\pi^+p\to pa_0^+\phi$ reaction.}
	\label{1}
\end{figure}

Within our approach, the production mechanism of the $\Delta(1940)$ and $\Delta(1920)$ resonances in the reaction $\pi^+p\to pa_0^+\phi$ consists of the standard $t-$ and $u-$channel as shown in Fig.~\ref{1}.

The production amplitude is calculated from the following effective Lagrangians~\cite{Nam:2019mon},
\begin{equation}
	\begin{split}
		\mathcal{L}_{\rho\phi\pi}=&\dfrac{g_{\rho\phi\pi}}{\sqrt{2}}\epsilon^{\mu\nu\alpha\beta}\partial_\mu\rho_\nu\partial_\alpha\phi_\beta,\\
		\mathcal{L}_{\phi NN}=&-g_{\phi NN}\bar{N}\left[\gamma^\mu-\dfrac{\kappa_{\phi NN}}{2m_N}\sigma_{\mu\nu}\partial^\nu\right]\phi_\mu N+h.c.,
	\end{split}
\end{equation}
for $\Delta(1920)$,
\begin{equation}
	\begin{split}
		\mathcal{L}_{\Delta^*N\pi}^{\frac{3}{2}^+}=&\dfrac{g_{\Delta^*N\pi}}{m_\pi}\bar{\Delta}^{*\mu}\Theta_{\mu\nu}(Z)\partial^\nu\tau\cdot\pi N+h.c.,\\
		\mathcal{L}_{\Delta^*Na_0}^{\frac{3}{2}^+}=&-\dfrac{ig_{\Delta^*Na_0}}{m_{a_0}}\bar{N}\partial^\mu\tau\cdot a_0\gamma_5\Theta_{\mu\nu}(Z)\Delta^{*\nu}+h.c.,
	\end{split}
\end{equation}
and for $\Delta(1940)$,
\begin{equation}
	\begin{split}
		\mathcal{L}_{\Delta^*N\rho}^{\frac{3}{2}^-}=&-\bar{\Delta}^{*\mu}\left(\dfrac{g_{\Delta^*N\rho}^{(1)}}{2m_N}\gamma^\alpha+i\dfrac{g_{\Delta^*N\rho}^{(2)}}{4m_N^2}\partial_N^\alpha+i\dfrac{g_{\Delta^*N\rho}^{(3)}}{4m_N^2}\partial_\rho^\alpha\right)\\
		&\times(\partial^\rho_\alpha g_{\mu\nu}-\partial^\rho_\mu g_{\alpha\nu})\rho^\nu N+h.c.,\\
		\mathcal{L}_{\Delta^*N\pi}^{\frac{3}{2}^-}=&-\dfrac{ig_{\Delta^*N\pi}}{m_\pi}\bar{\Delta}^{*\mu}\Theta_{\mu\nu}(Z)\gamma_5\partial^\nu\tau\cdot\pi N+h.c.,\\
		\mathcal{L}_{\Delta^*Na_0}^{\frac{3}{2}^-}=&-\dfrac{g_{\Delta^*Na_0}}{m_{a_0}}\bar{N}\partial^\mu\tau\cdot a_0\Theta_{\mu\nu}(Z)\Delta^{*\nu}+h.c.,\\
	\end{split}
\end{equation}
with
\begin{equation}
	\Theta_{\mu\nu}(Z)=g_{\mu\nu}-(Z+\frac{1}{2})\gamma_\mu\gamma_\nu,
\end{equation}
where $\kappa_{\phi NN}=0.2$~\cite{Meissner:1997qt} is the anomalous magnetic moment. Here, we introduce the off-shell parameter $Z$ in effective Lagrangians to include the off-shell effect of high-spin particles~\cite{Penner:2002ma, Mizutani:1997sd, Benmerrouche:1989uc}. We take the value for $\phi\rho\pi$ coupling from Ref.~\cite{Fan:2019lwc}. Other constants are determined from the partial decay widths, given in Table.~\hyperlink{tab1}{1}. It should be noted that we use the average values of the branching ratios listed in PDG~\cite{ParticleDataGroup:2022pth}. In view of the fact that $g_{\Delta^*N\rho}^{(2)}$ and $g_{\Delta^*N\rho}^{(3)}$ in $\Delta^*N\rho$ interaction have never been strictly calculated, here we only calculate $g_{\Delta^*N\rho}^{(1)}$ by $\Delta^*\to N\rho$ decay and ignore the $g_{\Delta^*N\rho}^{(2)}$ and $g_{\Delta^*N\rho}^{(3)}$ terms~\cite{Wang:2018vlv}. Due to the masses of $\Delta(1940)$ and $\Delta(1920)$ are very close to $pa_0$ threshold, taking their finite widths into account is essential. We include the finite width effect by using the following formula as~\cite{Wang:2023lnb, Roca:2005nm}
\begin{equation}
	\begin{split}
		\Gamma_{\Delta^*\to Na_0}=&-\dfrac{1}{\pi}\int_{(m_{\Delta^*}-2\Gamma_{\Delta^*})^2}^{(m_{\Delta^*}+2\Gamma_{\Delta^*})^2}ds\Gamma_{\Delta^*\to Na_0}(\sqrt{s})\\
		&\Theta(\sqrt{s}-m_N-m_{a_0})\mathrm{Im}\left\{\dfrac{1}{s-m_{\Delta^*}^2+im_{\Delta^*}\Gamma_{\Delta^*}}\right\}.
	\end{split}
\end{equation}
\begin{table}[htbp]
	\renewcommand{\arraystretch}{1.8}
	\tabcolsep=1.6mm
	\captionsetup{justification=raggedright}
	\caption*{TABLE~I. Coupling constants used in this work.}
	\hypertarget{tab1}{}
	\begin{tabular}[b]{ccccc}
		State & \makecell{Width\\(MeV)} & \makecell{Decay\\channel} & \makecell{Branching ratio\\adopted} & $g^2/4\pi$\\
		$\Delta(1940)$ & 350 & $N\rho$ & 0.80  & $5.44$\\
		 &  & $N\pi$ & 0.105  & $1.71\times10^{-2}$\\
		 &  & $Na_0(980)$ & 0.02  & $0.13$\\
		$\Delta(1920)$ & 300 & $N\pi$ & 0.125  & $1.75\times10^{-3}$\\
		 &  & $Na_0(980)$ & 0.04  & $4.12$
	\end{tabular}
\end{table}

Since hadrons are not pointlike particles, it is necessary to consider a form factor at each vertex, which can parameterize the structure of the hadron. Here, we introduce the form factor for $\rho$ meson exchange as
\begin{equation}
	f_M(k_M^2)=\left(\dfrac{\Lambda_M^2-m_M^2}{\Lambda_M^2-k_M^2}\right)^2,
\end{equation}
where $k_M$ and $m_M$ denote the four-momentum and mass of exchanged meson, respectively. Here, we take $\Lambda_\rho=1.2$~GeV~\cite{Fan:2019lwc} for $\rho$ meson exchange.

For intermediate baryons, we introduce the form factor as
\begin{equation}
	f_B(q_B^2)=\left(\dfrac{\Lambda_B^4}{\Lambda_B^4+(q_B^2-m_B^2)^2}\right)^2,
\end{equation}
with $q_B$ and $m_B$ the four-momentum and mass of intermediate baryon, respectively. The cut-off parameter for intermediate baryons are taken as $\Lambda_{\Delta(1940)}=\Lambda_{\Delta(1940)}=\Lambda_p=1.5$~GeV.

The propagators for the exchanged particles are expressed as
\begin{equation}
	G_\rho^{\mu\nu}(k)=-i\left(\dfrac{g^{\mu\nu}-\frac{k^\mu k^\nu}{m_\rho^2}}{k^2-m_\rho^2}\right),
\end{equation}
for $\rho$ meson,
\begin{equation}
	G_p(k)=\dfrac{i(\slashed{k}+m_p)}{k^2-m_p^2},
\end{equation}
for $p$ baryon, and
\begin{equation}
	\begin{split}
		G^\frac{3}{2}(q)=&\dfrac{i(\slashed{q}+M_R)}{q^2-M_R^2+iM_R\Gamma_R}\left[g_{\mu\nu}-\dfrac{1}{3}\gamma_\mu\gamma_\nu\right.\\
		&\left.-\dfrac{1}{3M_R}(\gamma_\mu q_\nu-\gamma_\nu q_\mu)-\dfrac{2}{3M_R^2}q_\mu q_\nu\right],
	\end{split}
\end{equation}
for spin-3/2 baryons, where $k$ and $q$ are the four-momentum for exchanged particles; $M_R$ and $\Gamma_R$ are the mass and width of intermediate resonances.

With the ingredients presented above, the total scattering amplitudes of $\pi^+p\to pa_0^+\phi$ reaction can be written as
\begin{equation}
	\begin{split}
		\mathcal{M}_{\Delta(1940)}^\rho=&\dfrac{ig_{\Delta^*Na_0}g_{\Delta^*N\rho}^{(1)}g_{\rho\phi\pi}}{2m_{a_0}m_N}f_\rho(k_1)f_{\Delta^*}(q_1)\bar{u}(p_5,s_5)p_4^\mu\\
		&\times\Theta_{\mu\nu}(Z)G_{\Delta^*}^{\nu\rho}(q_1)\gamma_\sigma(k_{1\rho}G_\rho^{\sigma\tau}(k_1)-k_{1\sigma}G_\rho^{\rho\tau}(k_1))\\
		&\times u(p_2,s_2)\epsilon^{\omega\tau\eta\beta}k_{1\omega}p_{3\eta}\epsilon_{\phi\beta}(p_3),\\
		\mathcal{M}_{\Delta(1940)}^p=&-\dfrac{2ig_{\Delta^*Na_0}g_{\Delta^*N\pi}g_{\phi NN}}{m_{a_0}m_\pi}f_p(k_2)f_{\Delta^*}(q_1)\bar{u}(p_5,s_5)\\
		&\times p_4^\mu\Theta_{\mu\nu}(Z)G_{\Delta^*}^{\nu\rho}(q_1)\Theta_{\rho\sigma}(Z)\gamma_5p_1^\sigma G_p(k_2)\\
		&\times\left[\gamma^\omega+\dfrac{\kappa_{\phi NN}}{2m_N}(p_3^\omega-\slashed{p}_3\gamma^\omega)\right]\epsilon_{\phi\omega}(p_3)u(p_2,s_2),\\
		\mathcal{M}_{\Delta(1920)}^p=&\dfrac{2ig_{\Delta^*Na_0}g_{\Delta^*N\pi}g_{\phi NN}}{m_{a_0}m_\pi}f_p(k_2)f_{\Delta^*}(q_2)\bar{u}(p_5,s_5)\\
		&\times p_4^\mu\gamma_5\Theta_{\mu\nu}(Z)G_{\Delta^*}^{\nu\rho}(q_2)\Theta_{\rho\sigma}(Z)p_1^\sigma G_p(k_2)\\
		&\times\left[\gamma^\omega+\dfrac{\kappa_{\phi NN}}{2m_N}(p_3^\omega-\slashed{p}_3\gamma^\omega)\right]\epsilon_{\phi\omega}(p_3)u(p_2,s_2).
	\end{split}
\end{equation}
The $p_2$, $p_3$, $p_4$ and $p_5$ represent the four-momentums of the $p$, $\phi$, $a_0$ and $p$, respectively. $k_1$ and $k_2$ correspond to the four-momentum of exchanged particle in Fig.~\hyperlink{1a}{1(a)} and Fig.~\hyperlink{1b}{1(b)}, respectively. $q_1$ and $q_2$ have the same meaning as $k_1$, $k_2$, but for $\Delta(1940)$ and $\Delta(1920)$.

The differential and total cross sections for this reaction can be obtained through
\begin{equation}
	\begin{split}
		d\sigma=&\dfrac{(2\pi)^4}{4\sqrt{(p_1\cdot p_2)^2-m_p^4}}\left(\dfrac{1}{2}\sum|\mathcal{M}|^2\right)d\Phi_3\\
		=&\dfrac{1}{(2\pi)^4}\dfrac{1}{\sqrt{(p_1\cdot p_2)^2-m_p^4}}\dfrac{|\vec{p}_3||\vec{p}_5^*|}{32\sqrt{s}}\left(\dfrac{1}{2}\sum|\mathcal{M}|^2\right)\\
		&dm_{\Sigma\eta}d\Omega_5^*d\mathrm{cos}\theta_3,
	\end{split}
\end{equation}
where $p_1$, $p_2$ represent the four-momentum of the initial particles $\pi$, $p$ at total center-of-mass frame; $\vec{p}_5^*$ stands for the three-momentum of the $p$ baryon in the center-of-mass frame of $pa_0$ pair.

\section{results}
\begin{figure}[htpb]
	\centering
	\includegraphics[width=0.45\textwidth]{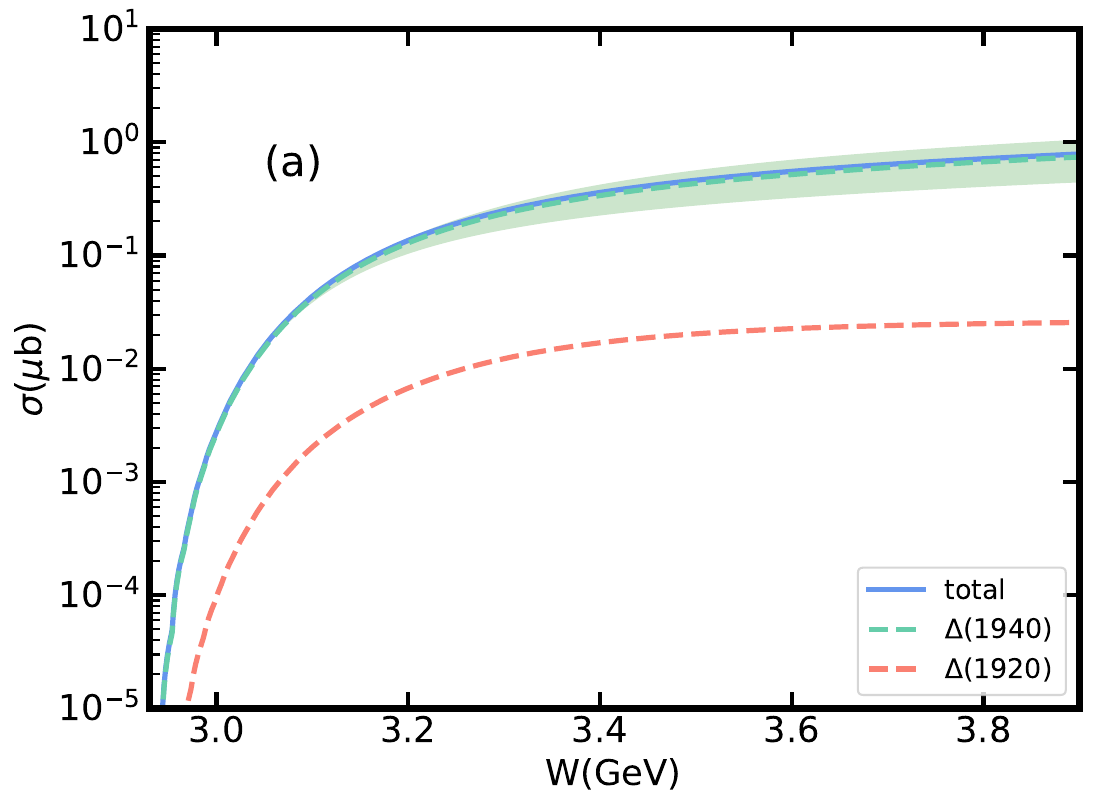}\hypertarget{2a}{}\\
	\includegraphics[width=0.45\textwidth]{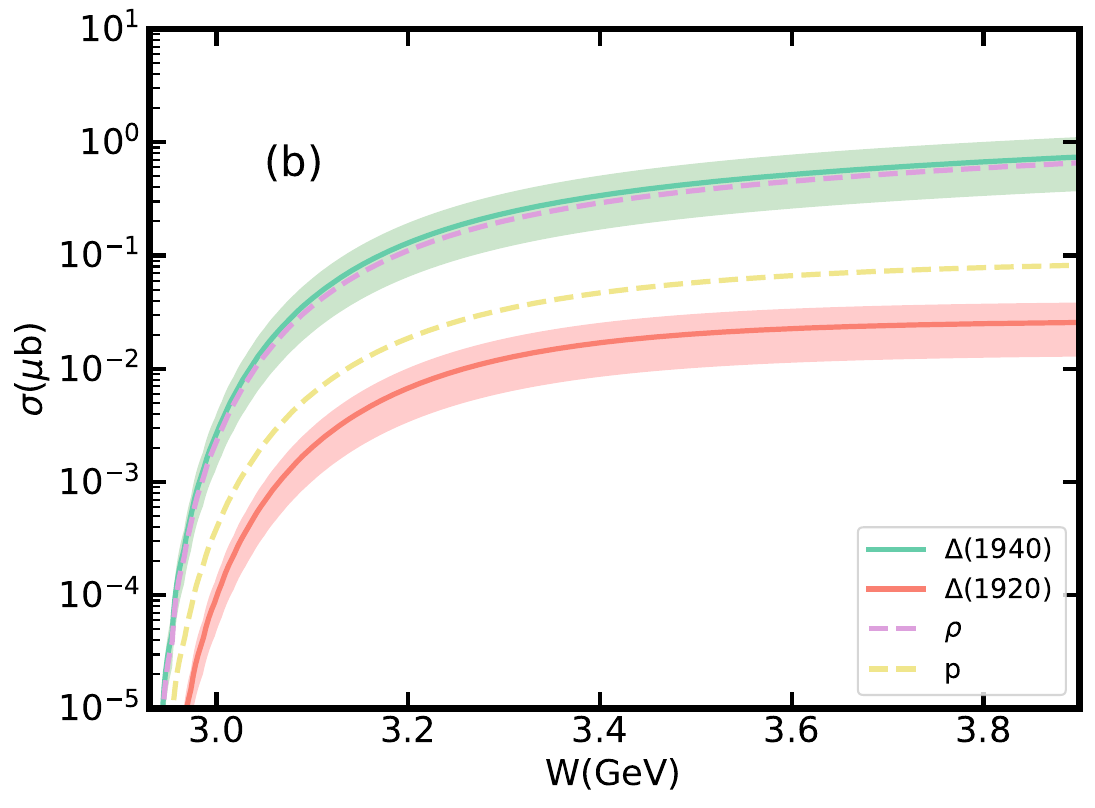}\hypertarget{2b}{}
	\captionsetup{justification=raggedright}
	\caption{Total cross section vs center of mass energy W for $\pi^+p\to pa_0^+\phi$ reaction. The blue curve represents the total cross section including all the contributions in Fig.~\ref{1}. The green band is the contribution of $\Delta(1940)$ with (a) cut-off parameter $\Lambda_{\Delta(1940)}$ varing from $1.2$ to $1.8$~GeV and (b) branching ratio for $\Delta(1940)\to pa_0$ decay varing from 1\% to 3\%. The red band is the contribution of $\Delta(1920)$ with branching ratio for $\Delta(1920)\to pa_0$ decay varing from 2\% to 6\%.}
	\label{2}
\end{figure}
In this section, we will present the theoretical results of the $\pi^+p\to pa_0^+\phi$ reaction calculated by the model in the previous section, including the total cross section and the differential cross section. Firstly, we display the predicted total cross section by fixing the cut-off parameters $\Lambda_{\Delta(1920)}=\Lambda_{\Delta(1940)}=1.5$~GeV. In this step, we ignore the off-shell effect, i.e., we set the off-shell parameter $Z=-0.5$, and the possible off-shell effects will be discussed later. In Fig.~\ref{2}, we plot the total cross section from the reaction threshold up to 3.9~GeV, together with the individual contributions of $\Delta(1940)$ and $\Delta(1920)$ resonances. For $\Delta(1940)$, both contributions of $\rho$ and $p$ exchanges are taken into account. For $\Delta(1920)$, only the contribution of $p$ exchange is considered since $\Delta(1920)N\rho$ coupling is not found in PDG. Obviously, $\Delta(1940)$ plays a dominant role of this reaction. As can be expected, on the one hand, that this is due to the relatively large $\Delta(1940)N\pi$ coupling. On the other hand, $\Delta(1920)$ lacks the coupling with $N\rho$ channel, and results in the absence of reaction process that shown in Fig.~\hyperlink{1a}{1(a)}, which further leads to an attenuated contribution of $\Delta(1920)$. In Fig.~\hyperlink{2b}{2(b)}, we present the cross sections from $\rho$ and $p$ exchanges, purple and yellow dotted lines respectively, compare to the $\Delta(1940)$ cross section. It is clear that even if only focus on the $p$ exchange, the contribution of $\Delta(1940)$ is still more significant than that of $\Delta(1920)$. This is due to the D-wave nature of $\Delta(1940)N\pi$ coupling and the large threshold momentum of this reaction, which enhance the $\Delta(1940)$ contribution compared to $\Delta(1920)$. Since $\Delta(1940)$ plays a dominant role near the threshold, it can be considered that this reaction provides a good place for studying the nature of $\Delta(1940)$ resonance. In our model, the final result contributed by $\Delta(1940)$ definitely still depends on the model parameters. The green band in Fig.~\hyperlink{2a}{2(a)} shows the uncertainty caused by the cut-off parameter $\Lambda_{\Delta(1940)}$ which varies from $1.2$ to $1.8$~GeV. One can see that $\Delta(1940)$ contribution is not sensitive to the variation of $\Lambda_{\Delta(1940)}$.  Considering a 20\% uncertainty of cut-off parameter, i.e., $\Lambda_{\Delta(1940)}=1.5\pm0.3$~GeV, we find that the uncertainty of the predicted cross sections near the threshold can reach to 0.53\%-1.46\%. In addition, we also consider the effect of branching ratios for $\Delta(1940)\to pa_0$ and $\Delta(1920)\to pa_0$ decay on cross sections, depicted in Fig.~\hyperlink{2b}{2(b)}, where the value of branching ratios will not affect the dominant role of $\Delta(1940)$. With $Br(\Delta(1940)\to pa_0)=2\pm1$\% and $Br(\Delta(1920)\to pa_0)=4\pm2$\%, the predicted uncertainties of total cross section near the threshold are 44.5\%-45.1\% and 5.1\%-5.9\%, respectively.

\begin{figure}[htpb]
	\centering
	\includegraphics[width=0.45\textwidth]{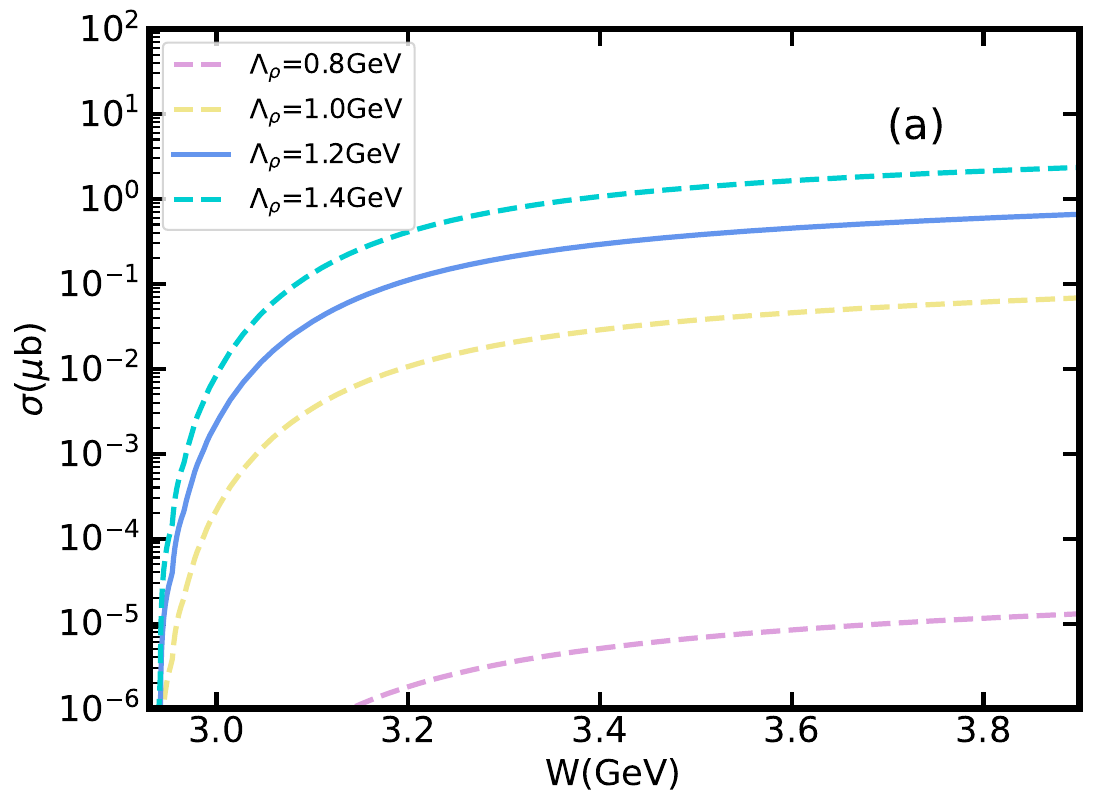}\hypertarget{3a}{}\\
	\includegraphics[width=0.45\textwidth]{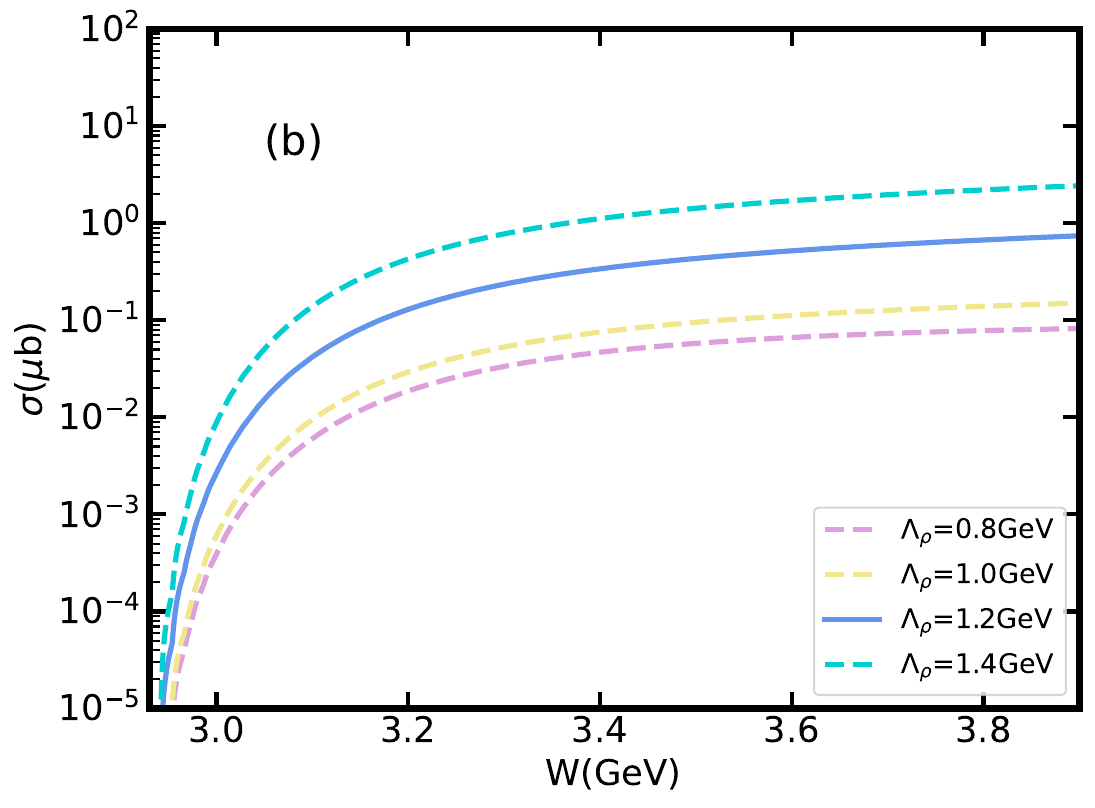}\hypertarget{3b}{}
	\captionsetup{justification=raggedright}
	\caption{The cross sections vs center of mass energy W from (a) the $p$ meson exchanged contribution and (b) the overall $\Delta(1940)$ contribution with different values of cut-off parameter $\Lambda_\rho$.}
	\label{3}
\end{figure}
Next, we consider the dependence of the cross section on the model parameter $\Lambda_\rho$ introduced by the form factor. The $\Lambda_{\Delta(1940)}$ cross sections from $\rho$ exchange with four cut-off parameters $\Lambda_\rho=0.8, 1.0, 1.2, 1.4$~GeV are displayed in Fig.~\hyperlink{3a}{3(a)}, where the $\rho$ exchange contribution has a strong dependence on cut-off parameter $\Lambda_\rho$. And the closer $\Lambda_\rho$ is to the mass of the $\rho$ meson, the stronger the dependence becomes. However, due to the effect of the larger $p$ exchange in the $\Delta(1940)$ resonance, the overall dependence of $\Delta(1940)$ contribution on $\Lambda_\rho$ is relatively small. The overall $\Delta(1940)$ cross sections with $\Lambda_\rho=0.8, 1.0, 1.2, 1.4$~GeV are presented in Fig.~\hyperlink{3b}{3(b)}. Since the cut-off parameter is often regarded as a free parameter, more experimental data are needed to determine it.

\begin{figure}[htpb]
	\centering
	\includegraphics[width=0.45\textwidth]{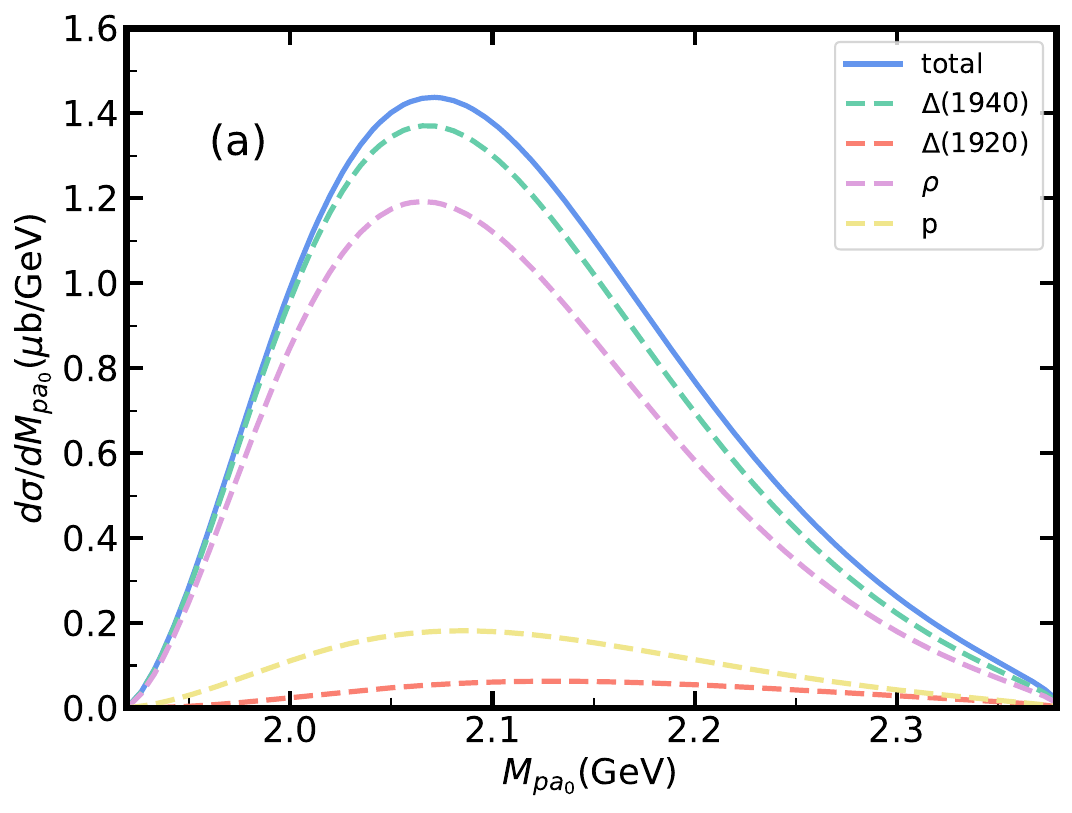}\hypertarget{4a}{}\\
	\includegraphics[width=0.45\textwidth]{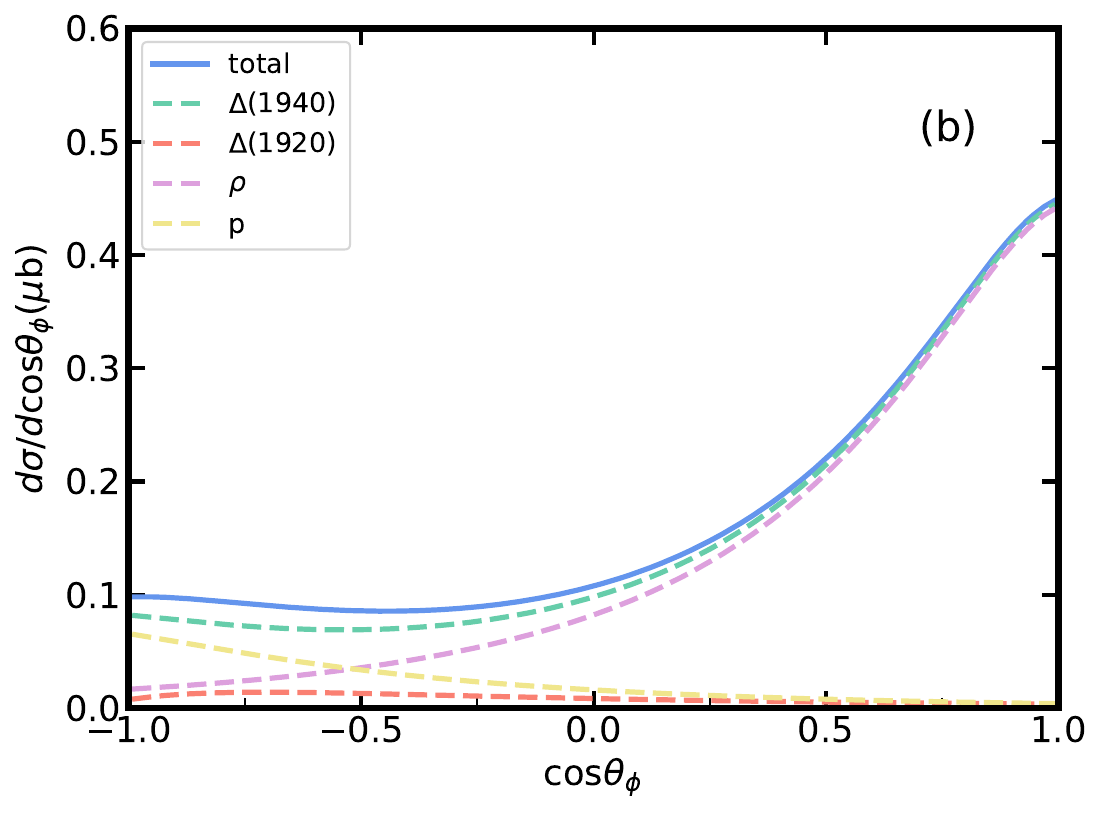}\hypertarget{4b}{}
	\captionsetup{justification=raggedright}
	\caption{At center of mass energy $W=3.4$~GeV, (a) invariant mass distribution of final $pa_0$ pair; (b) the angular distribution of $\phi$.}
	\label{4}
\end{figure}
In addition to the total cross section, we also study the differential cross section of $\pi^+p\to pa_0^+\phi$ reaction with the center of mass energy $W=3.4$~GeV, shown in Fig.~\ref{4}, where $\Delta(1940)$ is the main contribution. As can be seen from $pa_0$ invariant mass distribution shown in Fig.~\hyperlink{4a}{4(a)}, there is an obvious peak of $\Delta(1940)$ contribution. However, the shape of curve is relatively gentle due to the large width of $\Delta(1940)$ resonance. In Fig.~\hyperlink{4b}{4(b)}, on account of the comparatively large contribution of $\rho$ exchange, the total contribution is clearly presented as the forward enhancement of the angular distribution of $\phi$, and $p$ exchange in $u-$channel is manifested as an enhanced backward angle. Compared to $\Delta(1940)$, the contribution of $\Delta(1920)$ is so small in the angular distribution of $\phi$, and only slightly affects the shape of the total contribution in forward angle. 

\begin{figure}[htpb]
	\centering
	\includegraphics[width=0.45\textwidth]{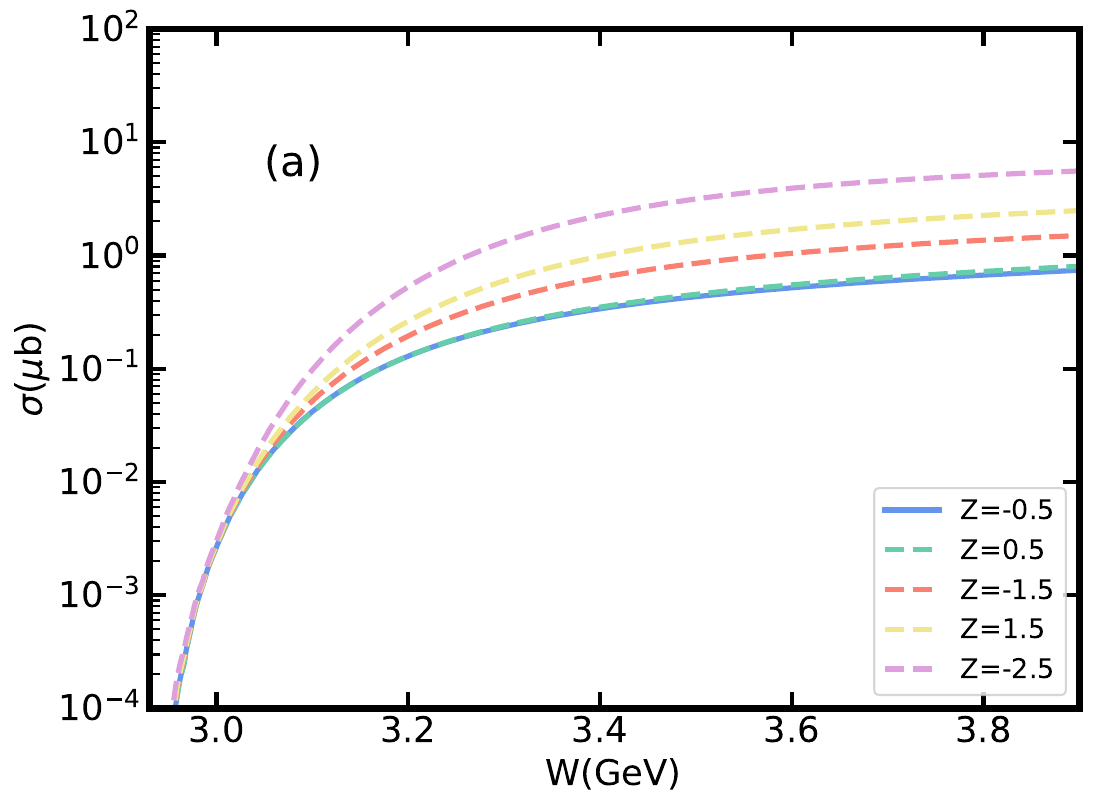}
	\includegraphics[width=0.45\textwidth]{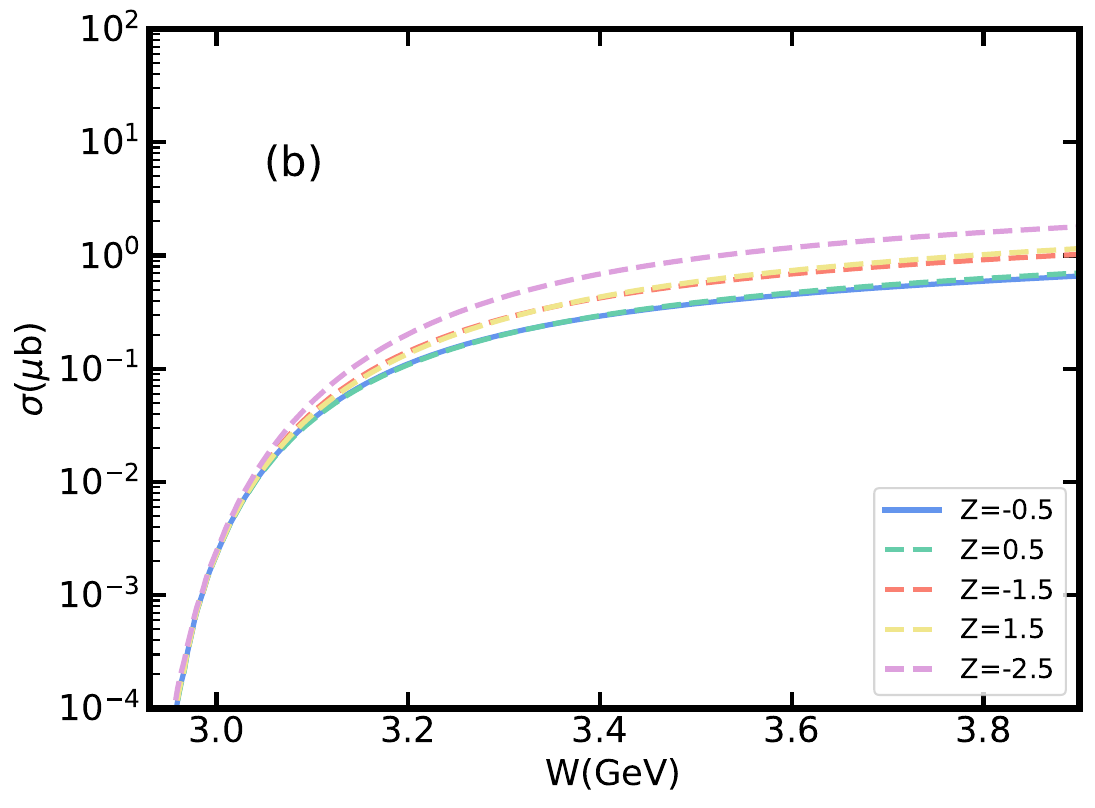}
	\includegraphics[width=0.45\textwidth]{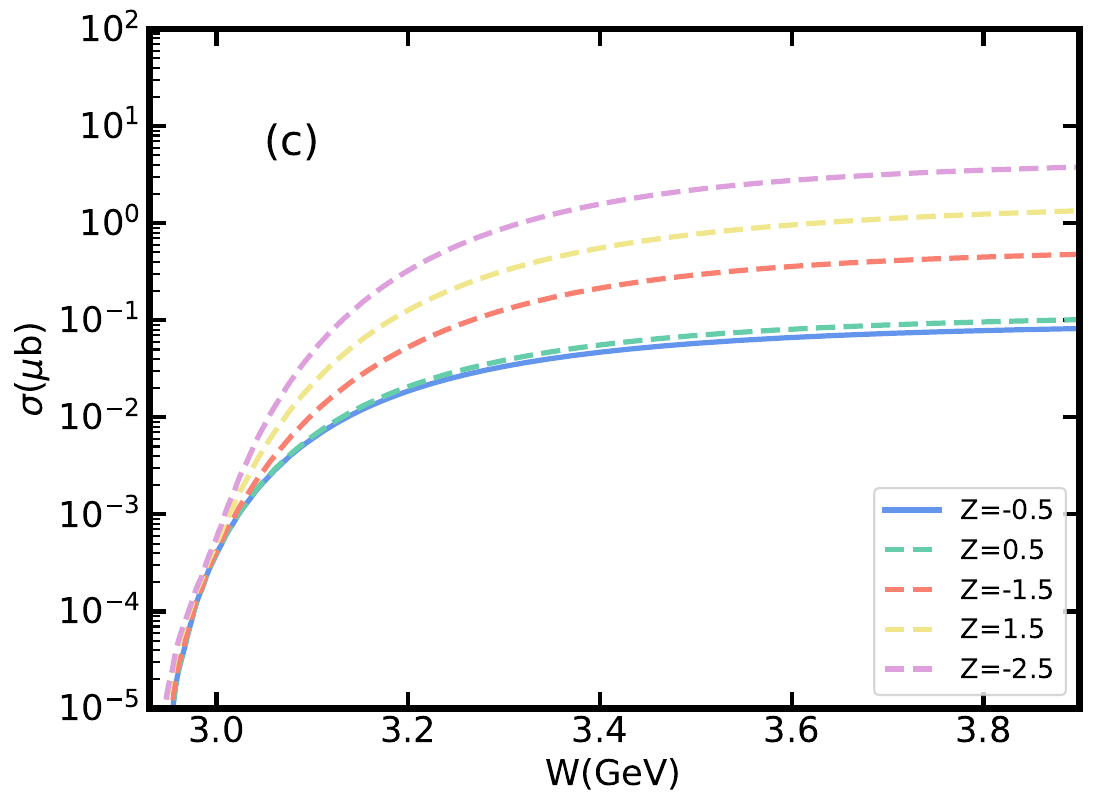}
	\captionsetup{justification=raggedright}
	\caption{The cross sections vs center of mass energy W with taking into account of the off-shell effects, where (a), (b) and (c) are the total cross section, $\rho$ exchange contribution and $p$ exchange contribution, respectively.}
	\label{5}
\end{figure}
Not only that, we also take into account the off-shell effects. First, we shall pay attention to the effect of the off-shell parameter $Z$ on $\Delta(1940)$ cross section. In Fig.~\ref{5}, we depict the cross sections with different off-shell parameters. As can be seen from the figures, the off-shell parameter have little effect on the contribution of $\rho$ exchange, and the variation of the total cross section shape is mainly due to the contribution of $p$ exchange. Since the effective Lagrangian of $\Delta N\rho$ vertex does not contain off-shell parameter, so that the amplitude of $\rho$ exchange contains only one off-shell parameter introduced by the effective Lagrangian of $\Delta Na_0$ vertex, making the cross section of $\rho$ exchange insensitive to variations in the off-shell parameter.

\begin{figure}[htpb]
	\centering
	\includegraphics[width=0.45\textwidth]{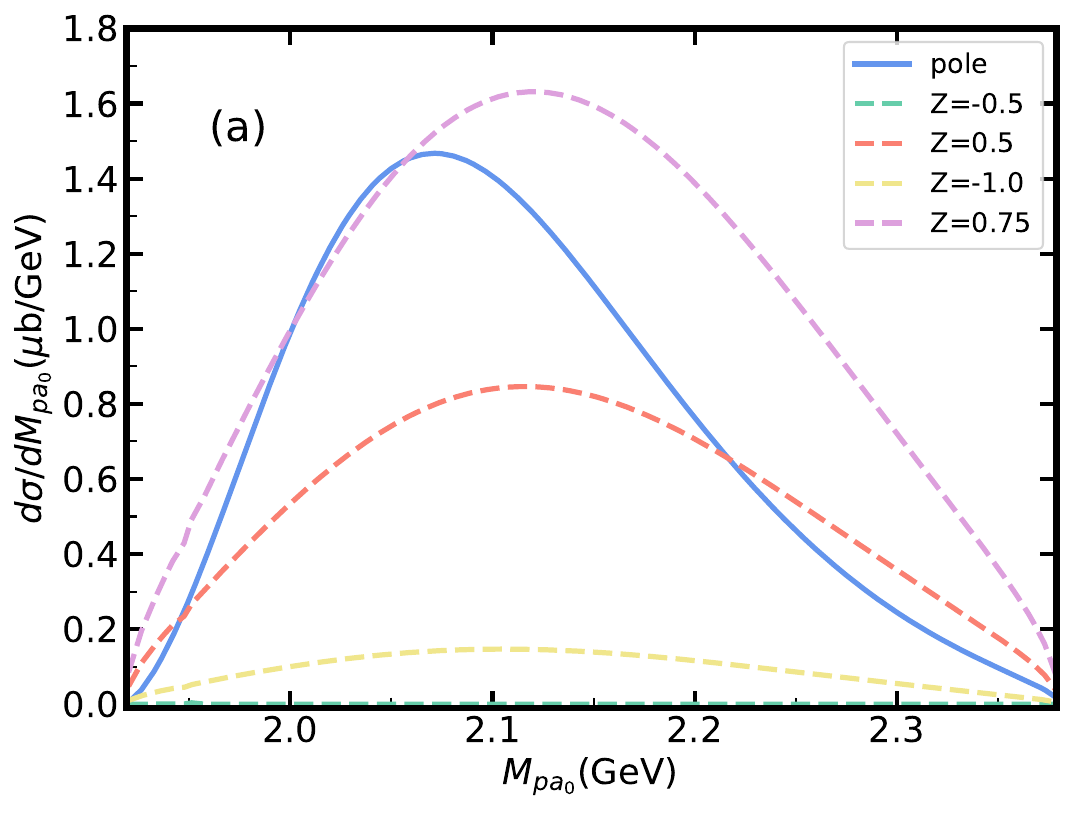}\hypertarget{6a}{}\\
	\includegraphics[width=0.45\textwidth]{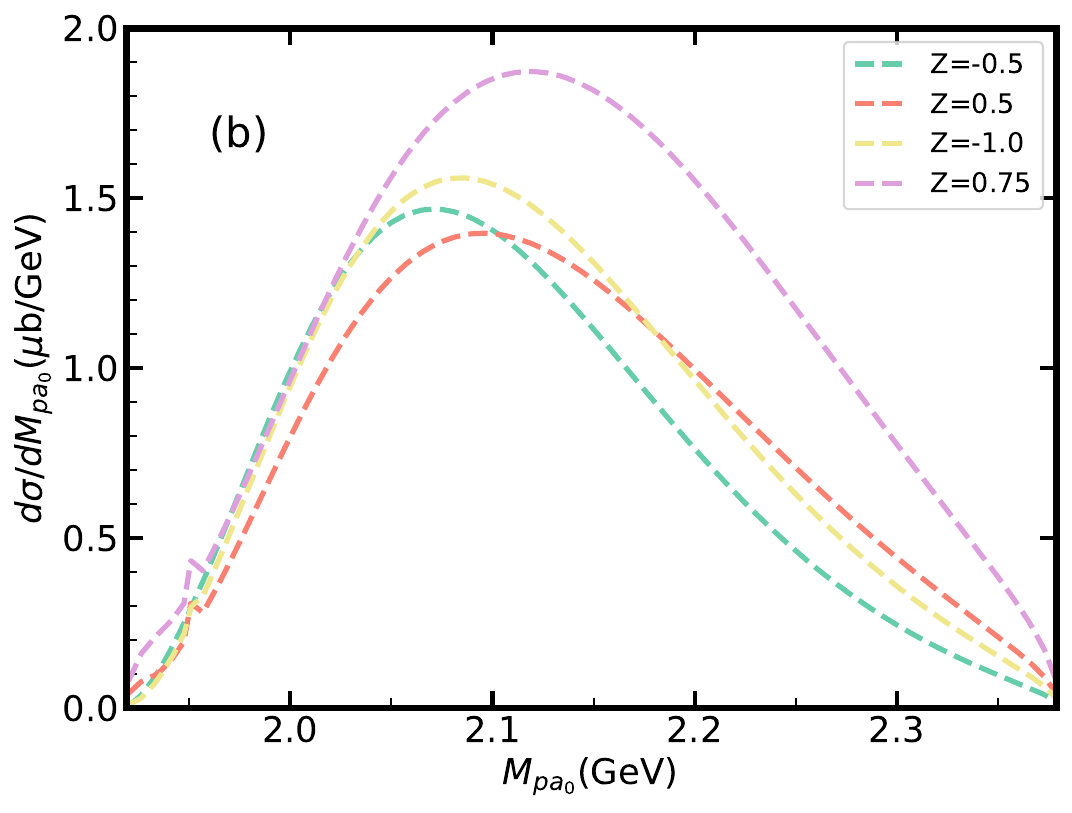}\hypertarget{6b}{}
	\captionsetup{justification=raggedright}
	\caption{At center of mass energy $W=3.4$~GeV, the invariant mass distribution of final $pa_0$ pair with taking into account of the off-shell effects.}
	\label{6}
\end{figure}
Then we focus on $M_{pa_0}$ spectrum with off-shell effects. As the discussion in Refs.~\cite{Wang:2023lnb, Mizutani:1997sd}, we analyze the amplitude of $\Delta(1940)$ decomposed into pole and nonpole parts. The contribution of pole part is independent of the off-shell parameter $Z$, while the freedom and effect of the off-shell parameter will only come from the nonpole part. And we adopt the idea in Refs.~\cite{Wang:2023lnb, Mizutani:1997sd} that the corresponding nonpole part may not dominate the pole part. The pole and nonpole contributions of $\Delta(1940)$ are shown in Fig.~\hyperlink{6a}{6(a)}, and Fig.~\hyperlink{6b}{6(b)} displays the full results of $\Delta(1940)$ contribution. The shape of $pa_0$ invariant mass distribution will change with the variation of the off-shell parameter, but the pole part can still reflect the structure of $\Delta(1940)$, which is not affected by the off-shell parameter. Therefore, the appearance of a resonance peak in $pa_0$ invariant mass distribution is not affected by the off-shell parameter.

\begin{figure*}[htpb]
	\centering
	\includegraphics[width=0.4\textwidth]{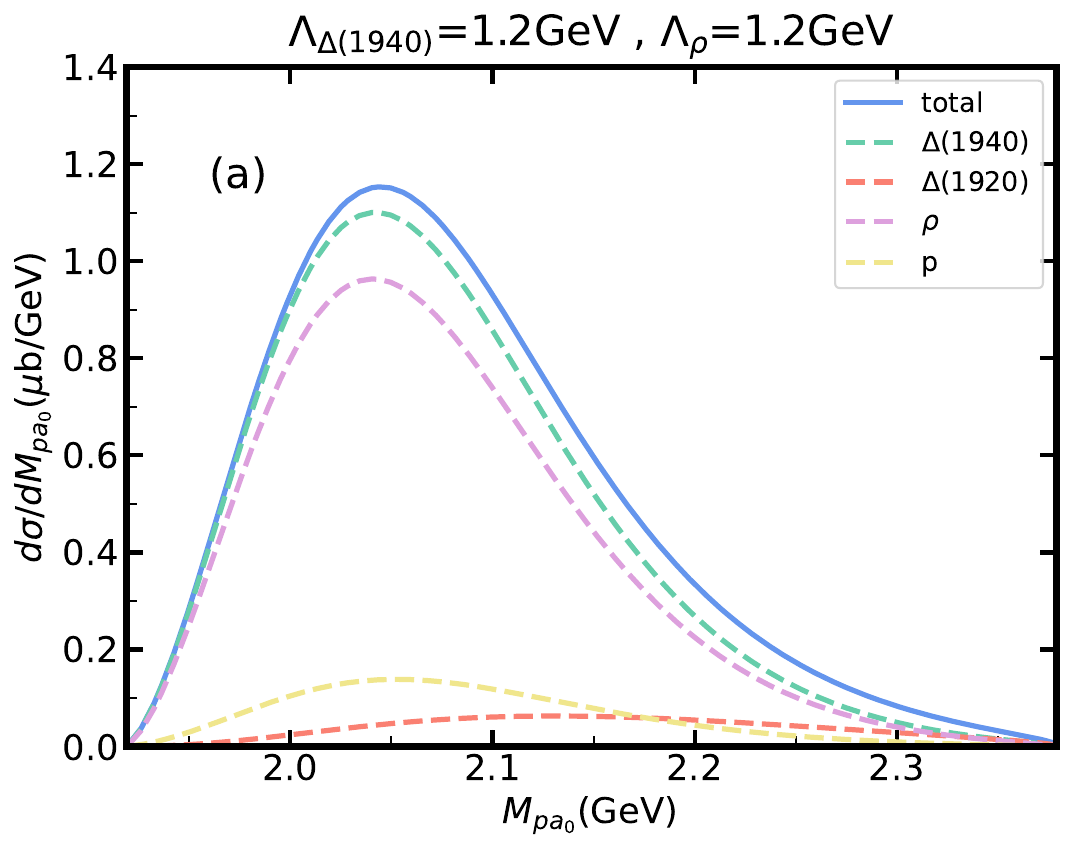}\hypertarget{7a}{}
	\includegraphics[width=0.4\textwidth]{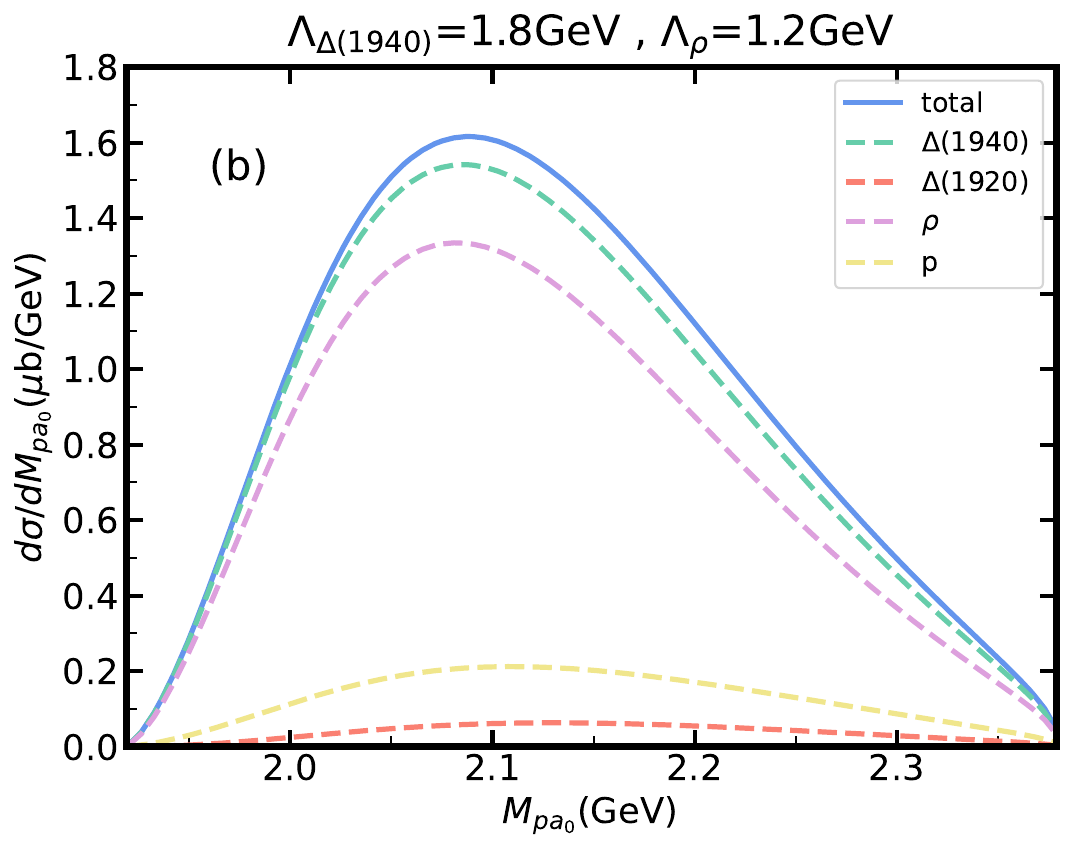}\\
	\includegraphics[width=0.4\textwidth]{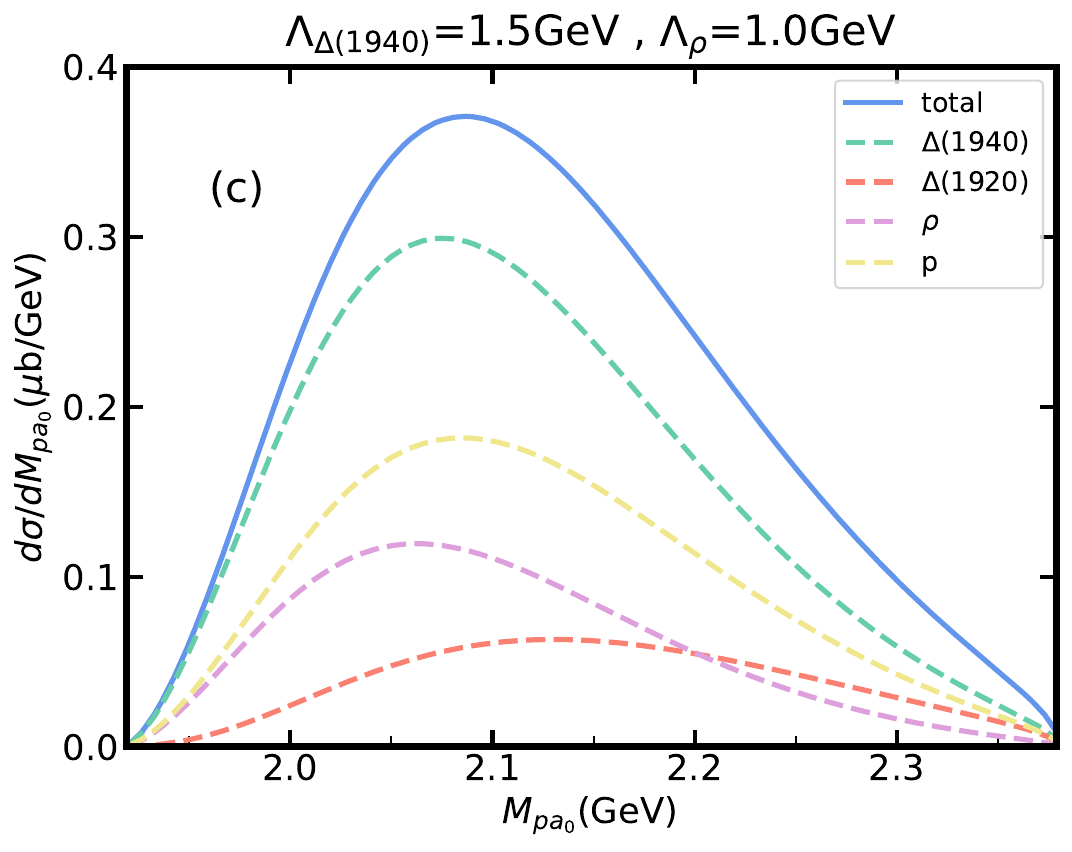}
	\includegraphics[width=0.4\textwidth]{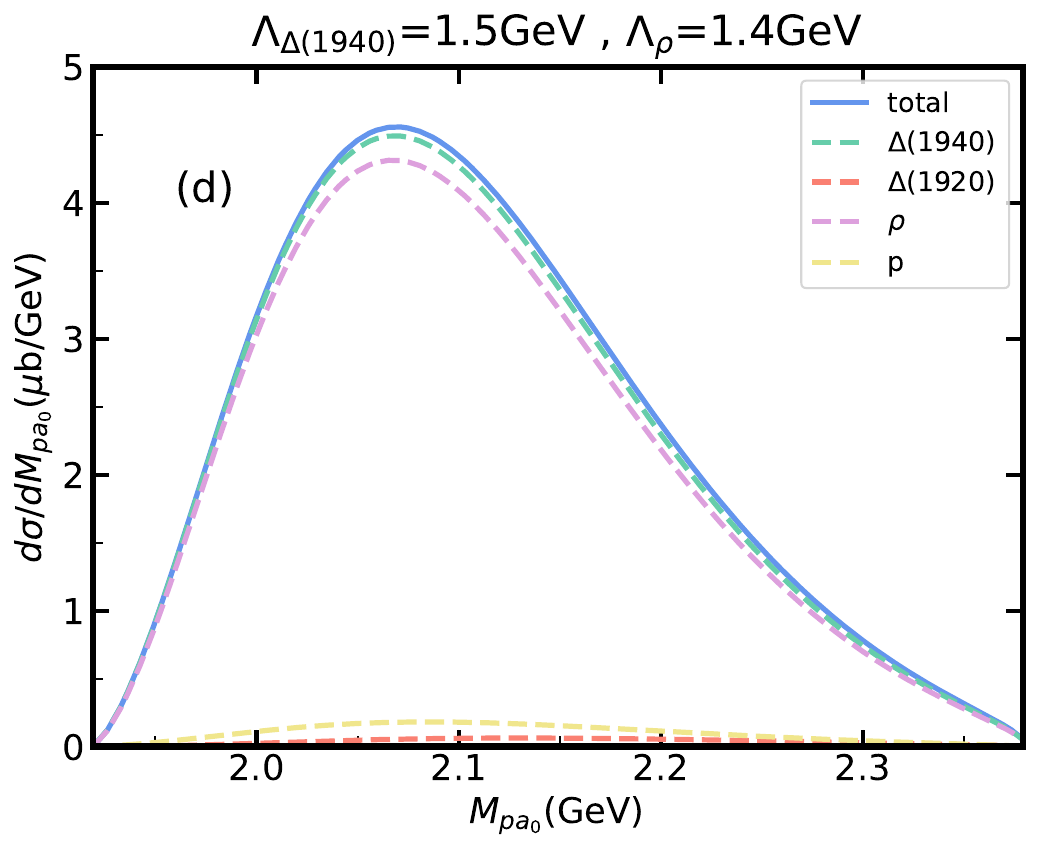}\\
	\includegraphics[width=0.4\textwidth]{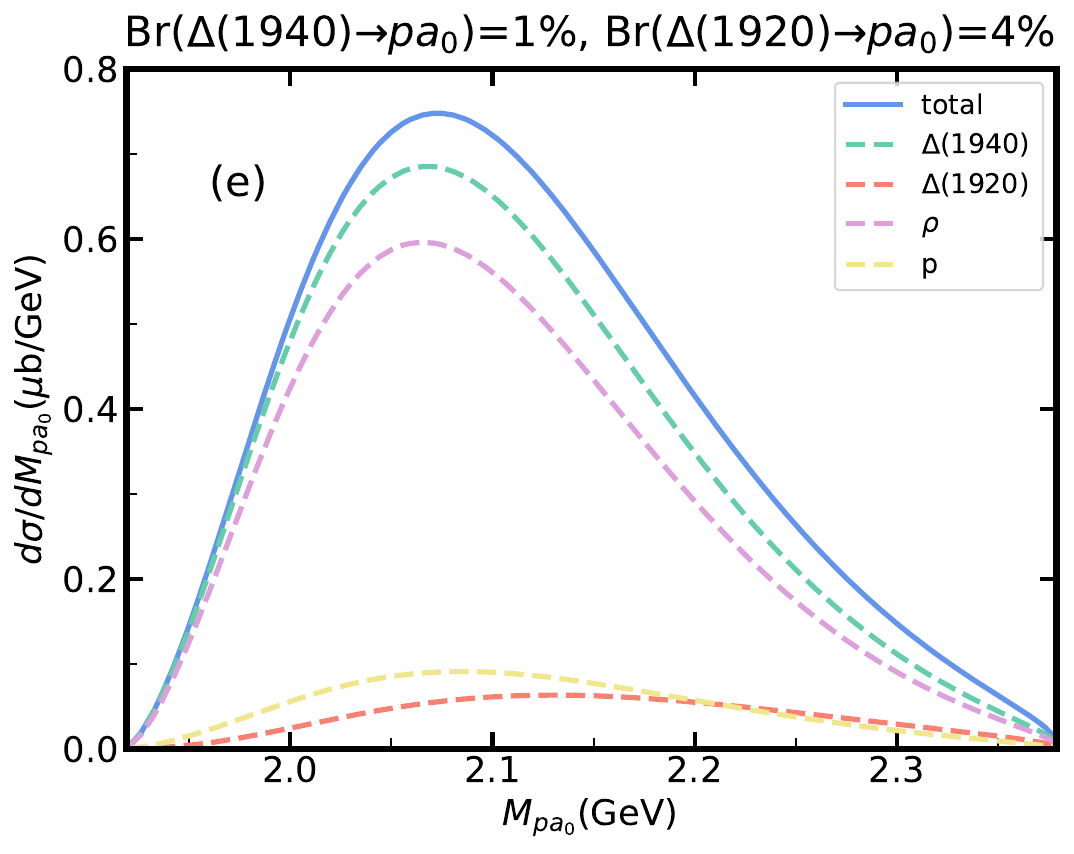}\hypertarget{7e}{}
	\includegraphics[width=0.4\textwidth]{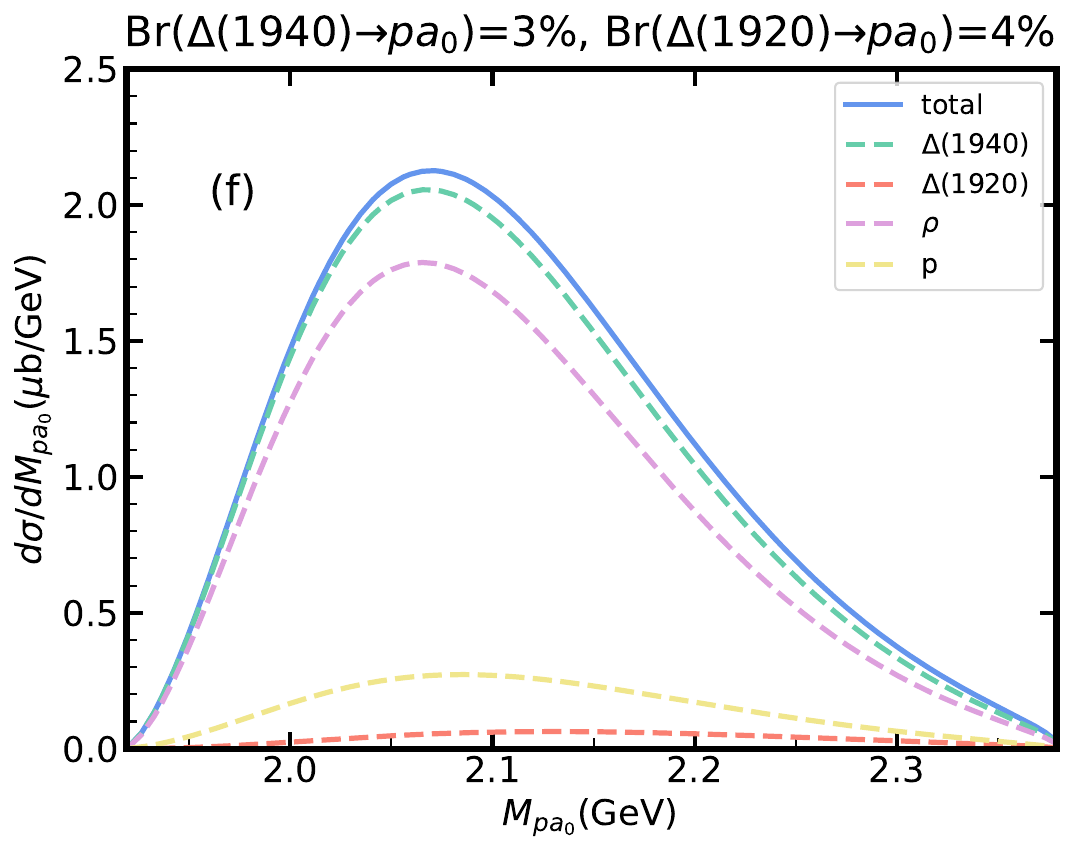}\\
	\includegraphics[width=0.4\textwidth]{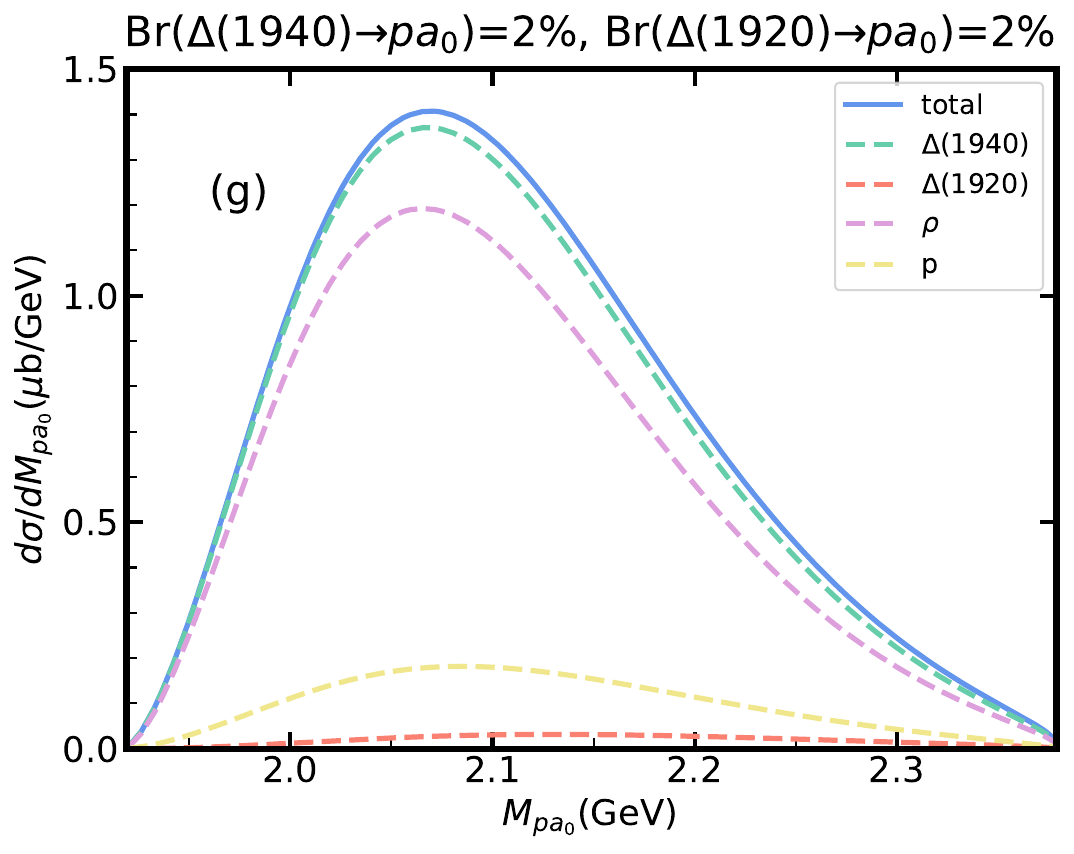}
	\includegraphics[width=0.4\textwidth]{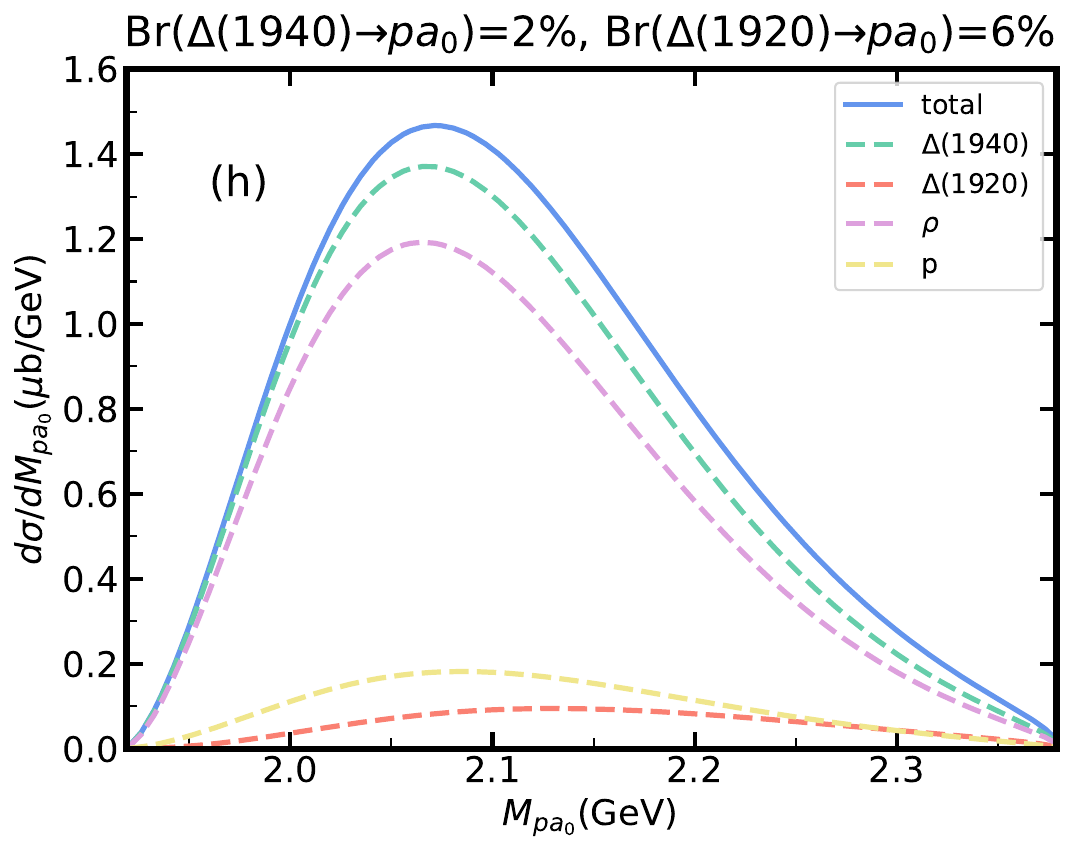}
	\captionsetup{justification=raggedright}
	\caption{At center of mass energy $W=3.4$~GeV, the invariant mass distribution of final $pa_0$ pair for $\pi^+p\to pa_0^+\phi$ reaction under different conditions, where (a)-(b) focus on the effect of $\Lambda_{\Delta(1940)}$; (c)-(d) focus on the effect of $\Lambda_\rho$.}
	\label{7}
\end{figure*}
\begin{figure}[htbp]
	\centering
	\includegraphics[width=0.45\textwidth]{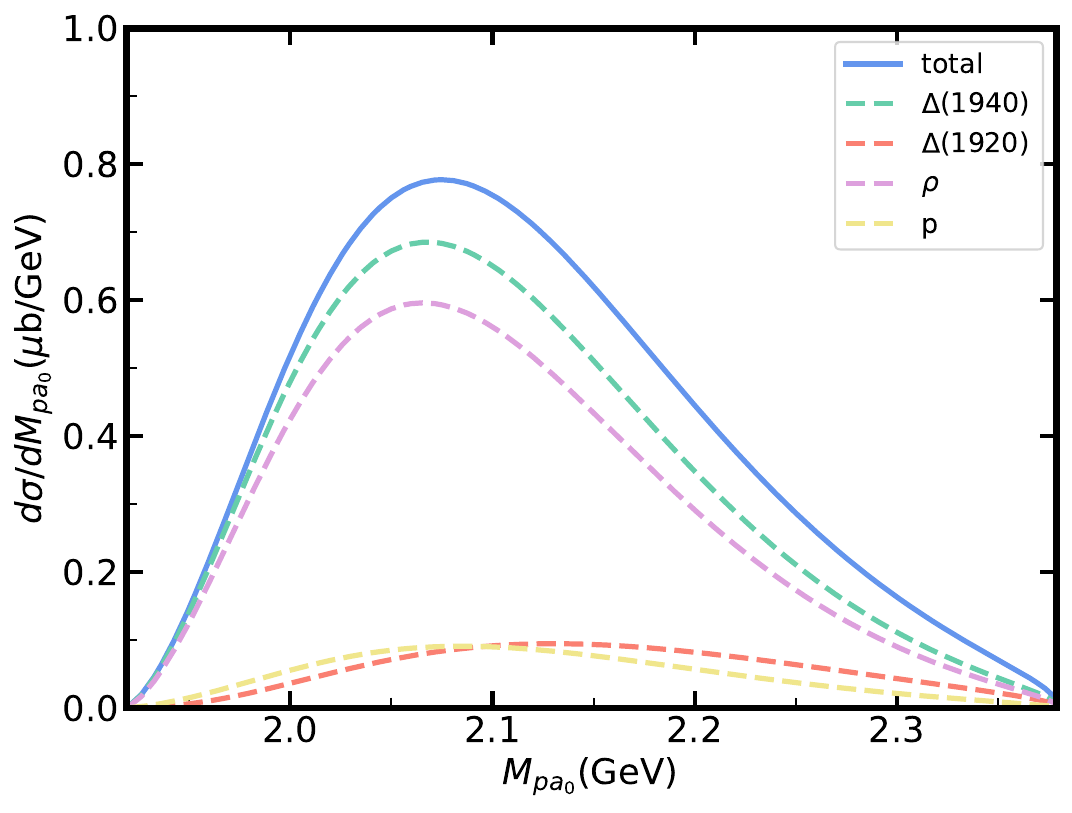}
	\captionsetup{justification=raggedright}
	\caption{At center of mass energy $W=3.4$~GeV, the invariant mass distribution of final $pa_0$ pair in the worst case with $Br(\Delta(1940)\to pa_0)=$1\% and $Br(\Delta(1920)\to pa_0)=$6\%.}
	\label{8}
\end{figure}
Similarly, the dependence of the peak position in $pa_0$ invariant mass distribution on the model parameters, $\Lambda_{\Delta(1940)}$ and $\Lambda_\rho$, also need to be considered. In Fig.~\hyperlink{7a}{7(a)-(d)}, we present the results of $pa_0$ invariant mass distribution by making $\Lambda_{\Delta(1940)}$ vary in 1.2, 1.8~GeV and $\Lambda_\rho$ vary in 1.0, 1.4~GeV, where the curve shape and peak position of the total contribution are slightly affected by cut-off parameters. However, when $\Lambda_\rho=1$~GeV, it will significantly reduce the contribution of $\rho$ exchange, which in turn reduces the $\Delta(1940)$ contribution. At this point $p$ exchange is the main contribution of $\Delta(1940)$. The effect of the branching ratios for $\Delta(1940)\to pa_0$ and $\Delta(1920)\to pa_0$ decays are shown in Fig.~\hyperlink{7e}{7(e)-(h)}, where $\rho$ exchange plays a dominant role in total contribution. Even we take $Br(\Delta(1940)\to pa_0)=$1\% and $Br(\Delta(1920)\to pa_0)=$6\%, i.e. Fig.~\ref{8}, due to the significant contribution of $\rho$ exchange, the total contribution is still completely dominated by the $\Delta(1940)$ resonance.
In $pa_0$ invariant mass distribution, although the peak position of the total contribution depends on the selection of cutoff parameters and decay branching ratio, the effect is not significant.

\section{conclusion}
We have made a theoretical study of $\pi^+p\to pa_0^+\phi$ reation baesd on an effective Lagrangian approach. In our model, we consider the productions of $\Delta(1940)$ and $\Delta(1920)$ as intermediate states excited by the $\rho$ and $p$ exchanges between the initial proton and $\pi$ meson. We provide a prediction of total and differential cross sections and discuss the possible influence of model parameters, off-shell effects and branching ratios for $\Delta(1940)\to pa_0$ and $\Delta(1920)\to pa_0$ decay. According to our results, $\Delta(1940)$ resonance makes a significant contribution near the threshold, making the reaction suitable for studying the features of $\Delta(1940)$ resonance and the coupling of $\Delta(1940)Na_0$ vertex. Both the cut-off parameters and branching ratios have a slight effect on the position of the peak in $pa_0$ invariant mass distribution. The peak position of the total contribution is very close to $\Delta(1940)$ contributes alone, even if we adopt $Br(\Delta(1940)\to pa_0)=$1\% and $Br(\Delta(1920)\to pa_0)=$6\%. And the off-shell effect will not affect the appearance of resonance peak. However, cut-off parameters, $\Lambda_{\Delta(1940)}$ and $\Lambda_\rho$, and  off-shell parameter $Z$ are usually regarded as free parameters, therefore, more experimental data are needed to determine them. Thus the measurements of this reaction will offer a good opportunity further explore the features of $\Delta(1940)$ resonance.

\section{acknowledgments}
HS is supported by the National Natural Science Foundation of China (Grant No.12075043). XL is supported by the National Natural Science Foundation of China under Grant No.12205002.

\bibliographystyle{unsrt}
\bibliography{c}
\end{document}